\documentclass[aps,pra,aps,twocolumn,a4paper,superscriptaddress,longbibliography,nofootinbib,showpacs]{revtex4-1}
\usepackage[english]{babel}
\usepackage{graphicx}
\usepackage{amsmath}
\usepackage{amssymb}
\usepackage{color}

\usepackage[colorlinks]{hyperref}

\usepackage[normalem]{ulem}   

\newcommand{\beq}{\begin{equation}}

\newcommand{\eeq}{\end{equation}}

\newcommand{\bra}[1]{\langle #1|}
\newcommand{\ket}[1]{|#1\rangle}

\newcommand{\md}[1]{\left|#1\right|}

\begin{document}

\title{Objectivity in non-Markovian spin-boson model}

\author{Aniello Lampo}\email{aniello.lampo@icfo.eu}
       \affiliation{ICFO -- Institut de Ci\`encies Fot\`o                                                                                                                                                        niques, The Barcelona Institute of Science and Technology, 08860 Castelldefels (Barcelona), Spain}
\author{Jan Tuziemski}
        
\affiliation{Faculty of Applied Physics and Mathematics, Technical University of Gda\'{n}sk, 80-233 Gda\'{n}sk, Poland}
\affiliation{National Quantum Information Center of Gda\'{n}sk, 81-824 Sopot, Poland}
\author{Maciej Lewenstein}
        \affiliation{ICFO -- Institut de Ci\`encies Fot\`o                                                                                                                                                        niques, The Barcelona Institute of Science and Technology, 08860 Castelldefels (Barcelona), Spain}
        \affiliation{ICREA, Psg. Lluis Companys 23, E-08010 Barcelona, Spain}
\author{Jaros\l aw K. Korbicz}
       
\affiliation{Faculty of Applied Physics and Mathematics, Technical University of Gda\'{n}sk, 80-233 Gda\'{n}sk, Poland}
 \affiliation{National Quantum Information Center of Gda\'{n}sk, 81-824 Sopot, Poland}


\begin{abstract}
Objectivity constitutes one of the main features of the macroscopic classical world. 
An important aspect of the quantum-to-classical transition issue is to explain how such a property arises from the microscopic quantum theory. 
Recently, within the framework of open quantum systems, there has been proposed such a mechanism in terms of the, so-called, spectrum broadcast structures.
These are multipartite quantum states of the system of interest and a part of its environment, assumed to be under an observation.
This approach requires a departure from the standard open quantum systems methods, as the environment cannot be completely neglected.
In the present paper we study the emergence of such a state structure in  one of the canonical models 
of the condensed-matter theory: the spin-boson model, describing the dynamics of a two-level system coupled to an environment made up by a large number of harmonic oscillators. 
We pay much attention to the behavior of the model in the non-Markovian regime, in order to provide a testbed to analyze how the non-Markovian nature of the evolution  affects the surfacing of a spectrum broadcast structure. 

 \end{abstract}
\date{\today}

\maketitle

\section{Introduction}
Quantum mechanics is one of the most successful theories, correctly predicting a huge class of physical phenomena. 
Its validity remains confined to the microscopic regime, where such a theory provides a good explanation of the behavior of the constituents of matter. In contrast, there is no trace of quantum effects on macroscopic scales, fully ruled by classical physics \cite{SchlosshauerBook, Schlosshauer2005, Zurek2003}. The rare counter examples  involve typically unstable systems, and phenomena such as super-radiance \cite{Altland2012}, superfluorescence \cite{Haake1979}, or spontaneous stimulated Raman scattering \cite{Walmsey1983}, where the quantum fluctuations might become macroscopically enhanced. 
Another prominent counterexample concerns obviously Bose-Einstein condensation and superfluidity and superconductivity \cite{Pita-String}. 
Despite these important but rather rare counterexamples, one natural question arises: How do the classical features of the macroscopic world emerge from the underlying quantum domain?

In particular, from everyday experience we are accustomed to perceive nature as \textit{objective}: 
We all observe the same (modulo different reference frames) properties of an observed object, without disturbing it.
This point of view has been fundamentally challenged by quantum mechanics, since the act of observation generally modifies a state of the observed system. 
So it is natural to wonder how 
the objective character of the classical theory can be derived from the (inherently subjective) quantum theory.

An important contribution to this fundamental problem has been given by the \textit{quantum Darwinism} program \cite{Zurek2009}, which is a more realistic and elaborated form of the decoherence theory.
It attributes objectification of information about a quantum system to an unavoidable interaction of the latter with its environment.
The main breakthrough of this approach lies in the role played by the environment: 
It is no longer a mere source of noise and dissipation, but is recognized to be an information carrier as indeed most of our everyday observations are indirect.
Moreover, the environment is considered to be divided into several different portions, $\{E_1,...,E_{N}\}$, each representing degrees of freedom available for observation for a single observer. Inevitably, some
portions will escape observation and are lost and hence, observed and unobserved portions of the environment deserve to be distinguished. 
A good example here is a visual observation of the same object by a group of people: Each person perceives a portion of the object's photonic environment, given by the solid angle of the observation.
But of course not all of the photons scattered by the object will be detected.
If it happens that each of these portions contains the same information about the object and it can be extracted without any disturbance, then we may speak of a certain, operational form of objective existence 
of this information  \cite{Zurek2009}.

One important step beyond the \textit{quantum Darwinism} program has been accomplished in \cite{Horodecki2015, Korbicz2014} where the authors analyzed so-defined operational objectivity directly in terms of quantum states. 
Under certain assumptions, they have proven that a state of the  system becomes objective if and only if a joint state of the system plus the observed portions of its environment, $\{E_1,...,E_{fN}\}$, 
is of the following form, called \textit{spectrum broadcast structure} (SBS):
\begin{eqnarray}\label{SBS}
&&\rho_{S:fE}=\sum_ip_i\ket{x_i}\bra{x_i}\otimes\rho^{E_1}_i...\otimes\rho^{E_{fN}}_i,\\
&&\rho^{E_k}_i\perp \rho^{E_k}_{i'} \textrm{ for every } i'\ne i \textrm{ and } k={1,\dots,fN}.
\end{eqnarray}
Here $\{\ket{x_i}\}$ is the so-called pointer basis of the central system to which it decoheres, $p_i$ are initial pointer probabilities,
and $\rho^{E_k}_i$ are some states of the observed parts of the environment with mutually orthogonal supports for different pointer index $i$. 
Simply put, the reason why this state structure corresponds to a form of objectivity is the following. The mutual orthogonality of the supports 
of $\rho^{E_k}_i$ means those states are one-shot perfectly distinguishable. Hence,   
by performing the right measurements (projections on the orthogonal supports of $\rho^{E_k}_i$), each of the observers extracts the same information about the system -- the 
value of the index $i$, enumerating the possible states the system can be in. On average (after forgetting the results) this extraction does not disturb neither the other observers nor the central system,
as the whole state in Eq.\ (\ref{SBS}) stays unchanged. The implication in the other direction -- that demanding objectivity in the above sense (under certain assumptions) uniquely leads
to the structure in Eq.\ (\ref{SBS}) is much more elaborate and can be found in \cite{Horodecki2015}. The central points of the reasoning are Bohr's definition of non-disturbance \cite{Wiseman} and, so called, 
strong objectivity: The only correlation between the portions of the environment is common information about the system. 
Let us stress the conceptual importance of the SBS states Eq.\ (\ref{SBS}): They put a rather philosophical notion of objectivity into a concrete form of a multipartite quantum state.
A form that in principle can be checked in concrete models.

Let us note that Eq.\ (\ref{SBS}) is an idealized structure
and an approach to it by real-life states has been analyzed in \cite{Mironowicz}. The quantities which control this approach are: i) the decoherence factor due to the unobserved part of the environment 
and ii) the fidelities of the states $\rho^{E_k}_i$ for different $i$'s  \cite{Korbicz2014}.
Spectrum broadcast structures have been so far found in a number of paradigmatic for the open quantum systems theory models, including: 
the illuminated sphere model \cite{Korbicz2014}, massive quantum Brownian motion \cite{Tuziemski2015,Tuziemski2015b,Tuziemski2016},
and the spin-spin model \cite{Mironowicz}. Recently it has also been shown that SBSs are typical for von Neumann measurements with measuring apparatuses
composed of a large number of degrees of freedom \cite{Korbicz2016}. This motivates further their studies as an important bridge in the quantum-to-classical transition.

The main purpose of the current paper is to investigate the objectification processes through the SBS formation in another canonical model of decoherence -- the spin-boson model. 
It consists of a two-level central system interacting with a large reservoir of bosonic modes \cite{Weiss, Leggett1987, SchlosshauerBook, BreuerBook, Ferialdi2017}. 
The model plays an important role in quantum computing, as well as in experiments on macroscopic quantum coherence, for instance in those aimed to analyze the role of quantum coherence in biological systems.  
An important part of the paper is devoted to explore the behavior of the model in the non-Markovian regime. 
By \textit{non-Markovianity} we mean the presence of memory effects making the evolution of the central system strongly dependent on its past history \cite{GardinerBook, BreuerBook}. 
This situation constitutes a rule rather than an exception, especially in the low-temperature regime, or when the interaction between the central system and the surrounding degrees of freedom gets sufficiently strong.
It is then a natural question to ask if and how the non-Markovianity affects objectification processes in this model. This analysis is one of the main goals of the present present. 
We would like to stress that there are many different definitions of non-Markovianity (see the recent review \cite{BreuerRev2016} or \cite{Addis2014}). 
Here we use the definition of \cite{Hall2014} based on non-positivity of decoherence rates.

The paper is organized as follows. 
In Sec.\ (\ref{SBMSec}) we introduce the spin-boson model. 
In Sec.\ (\ref{SBSSec}) we study the partially reduced state $\rho_{S:fE}$, which  describes the system plus a fraction of its environment.
We are focused on checking weather the partially reduced state approaches the SBS form in Eq.\ (\ref{SBS}),
which can be studied by calculating: (i) the decoherence factor induced by the unobserved part of the environment and  (ii) the fidelity  \cite{FG1999} for different values of the central spin of the fragments of the observed part of the environment.
Derivation of the fidelity as a function of the physical parameters of the model, such as the temperature, the coupling strength etc., is an original result of our paper.
In Sec.\ (\ref{NonMarkovianRegimeSec}) we move our analysis towards the non-Markovian regime. In particular, we look for the range of the model parameters defining the non-Markovian behavior. 
Such issues have been extensively studied before (see \cite{Addis2014,BreuerRev2016}), however our approach requires a distinction between the observed and the unobserved environment, which needs to be taken into account in the evaluation of non-Markovianity. 
This extension is also an original outcome of the present paper. 
Finally, in Sec.\ (\ref{DiscussionSec}) we focus on the analysis of an influence of the non-Markovianity on the emergence of a SBS. A similar problem has also  been treated in \cite{Galve2016} in the context of the quantum Brownian motion model. 
Based on our analysis, we conclude that there is no direct connection between the non-Markovianity and the objectification processes in the spin-boson model.
  

\section{Introduction to Spin-Boson Model}\label{SBMSec} 
The spin-boson model is described by the following Hamiltonian:
\begin{equation}
H=H_S+H_E+H_{int},\label{Ham}
\end{equation}
where:
\begin{equation}\label{FreeHamiltonians}
H_S=\frac{1}{2}\Omega\sigma_z,\quad H_E=\sum_i\left(\frac{p^2_i}{2m_i}+\frac{1}{2}m_i\omega^2_ix^2_i\right)
\end{equation}
are, respectively, the self-Hamiltonian of the central system and the environment. The former is represented by a two-level system while the latter is represented by a set of uncoupled harmonic oscillators. 
In Eq.\ (\ref{FreeHamiltonians}) we set $\hbar=1$, and hereafter we work with these units.  The interaction Hamiltonian is given by the expression:
\begin{equation}\label{IntHamSBM}
H_{int}=\sigma_z\otimes\sum_ig_i\left(a_i+a^\dagger_i\right).
\end{equation}

It is important to stress that the above model is not the most general one, since 
the interaction Hamiltonian in Eq.\ (\ref{IntHamSBM}) does not lead to dissipation processes because:
\begin{equation}\label{commute}
[H_S,\sigma_z]=0.
\end{equation}
In fact, in the literature this model is commonly called the \textit{pure dephasing spin-boson model}, although we still refer to it as \textit{spin-boson} for brevity.
In realistic systems the time scale for decoherence is typically many orders of magnitude shorter than the timescale of the energy exchange. Thus, our model can be regarded as a good representation 
of such rapid decoherence processes during which the energy dissipation is negligible. 

Because of Eq.\ (\ref{commute}), the self-Hamiltonian $H_S$ can be effectively neglected.
Passing to the interaction picture, the interaction Hamiltonian takes the following form:
	\begin{equation}
	H^{I}_{int}(t)=\sigma_z \otimes\left(g_k a^{\dagger}_k e^{i \omega_k t} +g_k a_k e^{-i \omega_k t}\right).  
	\end{equation}
	Since for two arbitrary instances of time $t,t'$, the commutator $[H_{int}(t),H_{int}(t')]$ is a \textit{c} number:
	\begin{equation}
	[H^{I}_{int}(t),H^{I}_{int}(t')] = -2i \sum_k |g_k|^2 \sin \left[ \omega_k(t-t') \right],
	\end{equation}
	the evolution can be easily solved, using e.g. the Magnus series: 
		
		\begin{eqnarray}\label{eq:evolution}
	U(t) =&&\exp\left(\frac{1}{2}\int_{0}^{t} dt' \int^{t'}_0 d t'' [H^{I}_{int}(t'),H^{I}_{int}(t'')]\right) \times\nonumber \\
	&&\exp\left(-i\int_0^{t} dt' H^{I}_{int}(t')  \right)\nonumber\\
	&&=e^{-i \xi(t)} \sum_{n=0}^{1} \ket{n} \bra{n}  \otimes \bigotimes_k \exp\left(-i\int_0^{t} dt' H^n_k(t')\right)  \nonumber \\
 \equiv&&e^{-i \xi(t)} \sum_{n=0}^{1} \ket{n} \bra{n}  \otimes \bigotimes_k U_{E_k}(n,t),
	\end{eqnarray}  
where $\xi(t)$ is a global phase factor, which we drop out as it is not important for our considerations,
	\begin{eqnarray}
	\label{eq:cham}
	H^n_k(t) \equiv (-1)^n \left(g_k a^{\dagger}_k e^{i \omega_k t} +g_k a_k e^{-i \omega_k t}\right),
	\end{eqnarray} 
and	$n=\pm 1$ are the eigenvalues of $\sigma_z$, 
while the evolution of the environment is governed by:  
	\begin{eqnarray}
	&&U_{E_k}(n,t) \equiv D_i\left([-1]^n\alpha_k(t) \right),\\ &&\alpha_i(t) =2\frac{g_k}{\omega_k}(1-e^{i \omega_k t}),
	\end{eqnarray}
	where $D(\alpha)$ is a displacement operator.

\section{Structure of a partially reduced state}
\label{SBSSec}
In this section we investigate the structure of a partially reduced state, describing the central spin and a fraction of its oscillatory environment. 
We derive tools that allow us to conclude if in the course of evolution the partially reduced state approaches the SBS form, defined in Eq.\ (\ref{SBS}).
We assume that the environment consists of $N$ oscillators, $fN$ ($0<f<1$)  out of which are under a potential observation  and constitute what we call the observed fraction $fE$ of the environment, while the rest 
is assumed to be lost and is the unobserved fraction $(1-f)E$. This promotion of a part of the environment from a mere source of noise  to an information carrier is the key point of the current approach,
inherited from the quantum Darwinism idea \cite{Zurek2009},
and a novelty in comparison with the traditional point of view in open quantum systems, 
where environment is always treated as a set of unobserved and uncontrollable degrees of freedom .

The partially reduced state is obtained by simply tracing out the non-observed fragment of the environment:
\begin{align}\label{SBS1}
\rho_{S:fE}(t)=\text{Tr}_{(1-f)E}\left[U(t)\rho_{0S}\otimes\bigotimes^N_{k=1}\rho_{0k}U(t)^{\dagger}\right],
\end{align} 
where as customary we assumed a fully product initial state for the global system:
\begin{equation}
\rho_{SE}(0)=\rho_{0S}\otimes\bigotimes^N_{k=1}\rho_{0k}.
\end{equation}
Eq.\ (\ref{SBS1}) can be expressed as follows:
\begin{align}
\rho_{S:fE}(t)=&\sum_{n}c^{nn}_{0S}\ket{n}\bra{n}\otimes\rho^{nn}_{f}(t)+\nonumber\\
+&\sum_{m} \sum_{n\neq m}\Gamma_{nm}(t)c^{nm}_{0S}\ket{n}\bra{m}\otimes\rho^{nm}_{f}(t),\label{SBS2}
\end{align} 
where
$
c^{nm}_{0S}\equiv\bra{n}\rho_{0S}\ket{m}
$,
and:
\begin{equation}
\rho^{nm}_{f}(t)\equiv\bigotimes^{fN}_{k=1}U_{E_k}(n,t)\rho_{0k}U_{E_k}(m,t)^{\dagger}\equiv\bigotimes^{fN}_{k=1}\rho^{(k)}_{nm}(t).
\end{equation}
The quantity:
\begin{equation}
\Gamma_{nm}(t)=\prod_{k\in(1-f)E}\text{Tr}\left[U_{E_k}(n,t)\rho_{0k}U_{E_k}(m,t)^{\dagger}\right]
\end{equation}
represents the decoherence factor between the state $\ket{n}$ and $\ket{m}$ of the central system due to the unobserved fraction 
of the environment $(1-f)E$. 

From Eq.\ (\ref{SBS2}) one sees that one necessary condition to approach an SBS is given by the usual decoherence condition: $\Gamma_{nm}(t)= 0$.
However, it is not sufficient. One has also to check whether the information deposited in the environment during the decoherence can be perfectly read out, i.e.,
if the system-dependent states of the fragments of the environment have non-overlapping supports \cite{Korbicz2014}: 
\begin{equation}\label{orthogonality}
\rho^{(k)}_{nn}(t)\rho^{(k)}_{mm}(t)= 0,
\end{equation} 
and hence are perfectly one-shot distinguishable. 
Among different measures of distinguishability, the most suitable turns out to be the generalized overlap  (also known as Uhlmann's fidelity \cite{Uhlmann1976}):
\begin{equation}
B(\rho_1,\rho_2)\equiv \text{Tr}\sqrt{\sqrt{\rho_1}\rho_2\sqrt{\rho_1}}. 
\end{equation}
One cannot expect Eq.\ (\ref{orthogonality}) to hold at the level of single fragments. To the contrary, since each of the unitaries $U_{E_k}(n;t)$ weakly depends on the parameter $n$, the states $\rho^{(k)}_{nn}(t) $ are almost identical for different $n$'s.
However, it can happen that by grouping subsystems of the observed part $fE$ into larger fractions, 
called macrofractions $\mathcal{M}$, one can approach the perfect distinguishability in Eq.\ (\ref{orthogonality}) at the level of
macrofraction states $\rho_{n}^{\mathcal{M}}(t)\equiv\bigotimes_{k\in \mathcal{M}}\rho^{(k)}_{nn}(t)$ \cite{Korbicz2014}. Generalized overlap is well suited for such tests 
due to its factorization with the tensor product: 
\begin{align}
B^{\mathcal{M}}_{nm}(t)&\equiv B\left[\rho_{n}^{\mathcal{M}}(t),\rho_{m}^{\mathcal{M}}(t)\right]\nonumber\\
&=\prod_{k\in\mathcal{M}}B\left[\rho^{(k)}_{nn}(t),\rho^{(k)}_{mm}(t)\right].\label{Fid0}
\end{align}
We stress that the measure of distinguishability we introduce refers to macrofractions of the observed part of the environment.

Summarizing, the formation of SBS in Eq.\ (\ref{SBS}) is equivalent to \cite{Mironowicz}:
\begin{equation}
|\Gamma_{nm}(t)|\approx 0,\quad B^{\mathcal{M}}_{nm}(t)\approx 0.
\end{equation}
We will refer to the decoherence factor and the fidelity as to indicator functions. 
As it is well known, for the thermal environment:
\begin{equation}\label{EnvironmentThermalIntialConditions}
\rho_{0E}=\frac{e^{-\beta H_E}}{Z_E}
\end{equation}
where $\beta=1/T$ and $Z_E$ is the partition function,
the decoherence factor can be written in the analytical form \cite{BreuerBook, SchlosshauerBook}:

\begin{equation}\label{dec}
\Gamma(t)=\exp\left[-2\int_{(1-f)E} d\omega \coth\left(\omega/2T\right)g(\omega,t)\right],
\end{equation}
where:
\begin{equation}
g(\omega,t)\equiv J(\omega)\frac{1-\cos\left(\omega t\right)}{\omega^2},
\end{equation}
and the integral is performed over the frequency range related to the unobserved part of the environment $(1-f)E$.
The quantity $J(\omega)$ is the spectral density containing the information of the coupling with the environment  \cite{BreuerBook, SchlosshauerBook}:
\begin{equation}
J(\omega)=2\sum_{i}\md{g_i}^2\delta\left(\omega-\omega_i\right).
\end{equation} 
We postpone to next Section further discussion of the spectral density as it plays a fundamental role in defining Markovian and non-Markovian behavior.  

The expression for the fidelity in the present model constitutes in turn the result:
\begin{equation}\label{fid}
B(t)=\exp\left[-2\int_{\mathcal{M}} d\omega \tanh\left(\omega/2T\right)g(\omega,t)\right],
\end{equation}
where the integral is performed over the frequency range associated with a macrofraction $\mathcal{M}$.
The derivation is presented in Appendix \ref{FidelitySec} for the case of the initial thermal state of the environment (see \cite{Tuziemski2015b}).  
The most important difference between the fidelity  Eq.\ (\ref{fid}) and the decoherence  Eq.\ (\ref{dec}) lies in the dependence on the temperature. 
In particular, when $T\rightarrow0$ both quantities approach the same form, and the emergence of a SBS  reduces just to decoherence. 
This is in agreement with the fact that at low temperature the state of the fragments of the environment becomes pure. 
 
\section{Non-Markovian environments}\label{NonMarkovianRegimeSec}
One of the main purposes of the current paper is to investigate if and how non-Markovianity affects the emergence of objective properties in the pure-dephasing spin-boson model.
As presented in the introduction, the technical tools we use for checking for objectification are SBS states (\ref{SBS}) -- their formation implies objectification of a state of the central spin as it becomes redundantly stored in the bosonic environment
in many perfect copies.
Yet more technically, we argued in the previous section that the above state structures arise when both the decoherence factor and the appropriate fidelity vanish. 
These quantities, as well as the amount of non-Markovianity,  depend on the system parameters: the temperature and the couplings as summarized by the spectral density. 
We proceed thus by manipulating appropriately these parameters in order to increase non-Markovianity, and to analyze how decoherence and fidelity react.

First, we have to carefully introduce a measure of non-Markovianity, adapted to our specific scenario.
We will use the measure from \cite{Hall2014} but based on the dynamics of the partially reduced state, i.e. the state of the central spin and the observed part of the environment. 
The evolution of the latter [see (\ref{SBS2})] can be expressed in terms of  the following  local-in-time master equation (see Appendix \ref{app:canlk} for the details):
\begin{eqnarray}
\label{eq:canlk}
\dot{\rho}_{S:fE} = &&-i\left[H_{S:fE}(t),\rho_{S:fE}\right]\\ && + \gamma(t) \left( \sigma_z \otimes I_{fE} \rho_{S:fE} \sigma_z \otimes I_{fE} - \rho_{S:fE}\right), \nonumber
\end{eqnarray}
where:
\begin{eqnarray}
&&H_{S:fE}(t) \equiv \sum_n  \ket{n}\bra{n} \otimes H^n_{fE}(t), \; H^n_{fE}(t) \equiv \bigotimes_{k=1}^{fN} H^n_k(t) \nonumber \\
\end{eqnarray}	 and
\begin{equation}
\label{gamma}
\gamma(t) = \frac{\dot{\Gamma}(t)}{\Gamma(t)}.
\end{equation} 
The structure of Eq.\ (\ref{eq:canlk}) allows one to identify $\gamma(t)$ as the only non-zero canonical decoherence rate for the considered problem \cite{Hall2014}: Its positivity guarantees that the evolution is completely positive or ``time-dependent Markovian" \cite{Rivas2010,Hall2014}.
Accordingly, if $\gamma(t)$ becomes strictly negative the evolution is non-Markovian. 
To quantify the amount of non-Markovianity in a given interval of time we will adopt the measure introduced in \cite{Hall2014}: \begin{equation}
\label{NonMarkovMeasure}
\mathcal{N}=-\int_{\gamma<0}\gamma(t)dt.
\end{equation}
The above expression is formally similar to definitions of non-Markovianity based on the reduced dynamics of the spin only \cite{Rivas2010,Addis2014,Hall2014}. However, as we are interested in the structure of the partially reduced state,  in the present case $\gamma(t)$ incorporates non-unitary effects caused by the non-observed fraction of the environment. To the best of our knowledge such a situation has not been studied so far in the context of non-Markovianity measures.

In the high-temperature limit, $T \gg s \Lambda$, Eq.\ (\ref{NonMarkovMeasure}) becomes:
\begin{equation}\label{gammaT}
\gamma(t)=T\int_{(1-f)E} d\omega\frac{J(\omega)}{\omega^2}\sin\left(\omega t\right). 
\end{equation}
Let us now specify the spectral density. We will follow the usual approach and choose \cite{SchlosshauerBook,BreuerBook}:
\begin{equation}\label{SD}
J(\omega)=\frac{\omega^s}{\Lambda^{s-1}}\exp\left(-\omega/\Lambda\right). 
\end{equation}
where $\Lambda$ is the cutoff frequency and $s$ is the Ohmicity parameter. 
The case $s=1$ corresponds to pure Ohmic, whereas $s<1$ to sub- and $s>1$ corresponds to super-Ohmic regimes. 
The spectral density provides an information about the coupling of the two-level system with the environment. 
We note that he integral in Eq.\ (\ref{gammaT}) is performed over the frequencies belonging to the unobserved part of the environment. 
We will use the spectral density in Eq.\ (\ref{SD}) in two basic ways to define the observed and unobserved parts of the environment: 
(i) Each part is so large that the full spectral density applies to it (we call it uncut 
); and 
(ii) Each part is defined by some frequency range of Eq.\ (\ref{SD}) only (cut case).

\begin{figure}
\begin{center}
\includegraphics[width=0.8\columnwidth]{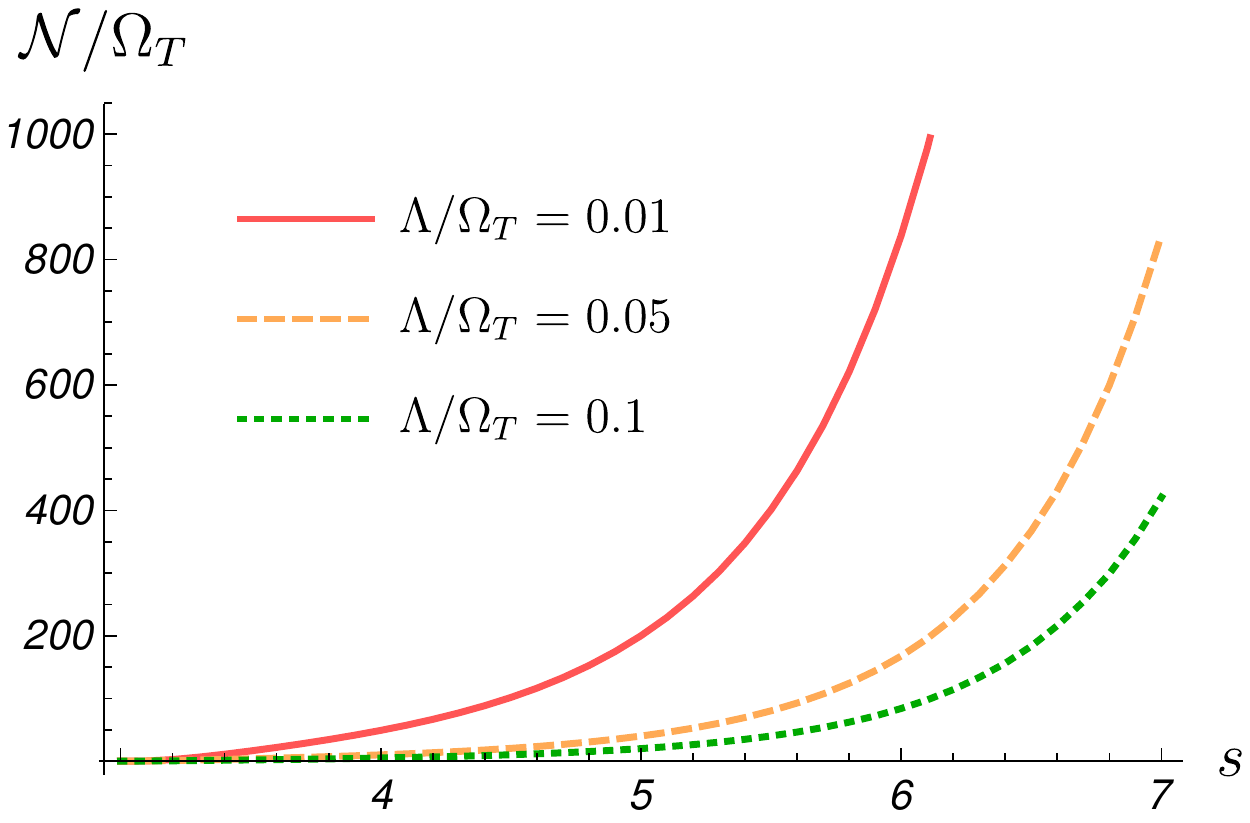}
\caption{\label{nSPlot} Non-Markovianity measure in Eq.\ (\ref{NonMarkovMeasure}) as a function of the Ohmicity parameter.}
\end{center}
\end{figure} 

Let us first briefly recall known facts about the connection between the non-Markovianity and the uncut spectral density (see e.g. \cite{Haikka2013}).
Fig.\ \ref{nSPlot} shows the plot of the measure in Eq.\ (\ref{NonMarkovMeasure}) as a function of the Ohmicity parameter $s$  for different values of the cutoff $\Lambda$ scaled to the thermal frequency $\Omega_T=k_BT/\hbar$.
The degree of non-Markovianity monotonically increases with $s$ for $s>3$, as for $s<3$ the dynamics is purely Markovian, i.e. $\mathcal{N}=0$. 
Moreover, in \cite{Addis2014}, where the measure in Eq.\ (\ref{NonMarkovMeasure}) has been calculated in the zero-temperature limit, it has been proved that the threshold at $s=3$, distinguishing the Markovian regime from the non-Markovian one, holds even for other measures. 

Fig.\ \ref{nSPlot} also provides the dependence on the cutoff $\Lambda$ of the present measure, at fixed $s$. Precisely, it monotonically decreases as $\Lambda$ grows, whatever values of the Ohmicity parameter we consider. This result could also be inferred by looking to the self-correlation function for the environment, decaying exponentially as $1/\Lambda$, as discussed in \cite{BreuerBook}. Finally, it is easy to infer, considering Eq.\ (\ref{gammaT}), that in the high-temperature limit non-Markovianity decreases linearly as $T$ grows, provided $s>3$. 

\begin{figure}
\begin{center}
\includegraphics[width=0.8\columnwidth]{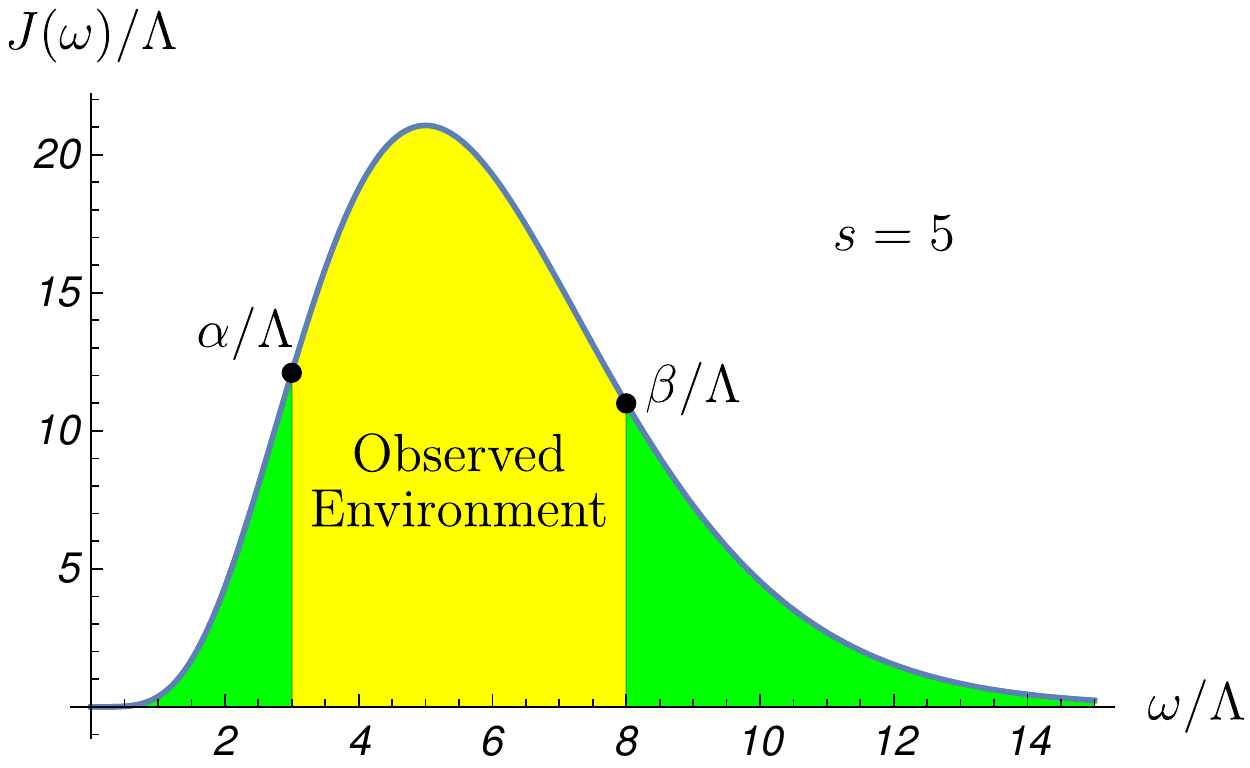}
\caption{\label{SDCuts2} Spectral density in Eq.\ (\ref{SD}). Observed and unobserved part of environment are represented, respectively, by the yellow and green portion of spectral density.}
\end{center}
\end{figure}

So far we have described the unobserved part of the environment by means of the whole spectral density in Eq.\ (\ref{SD}). Now, we consider a more realistic case in which such a spectral density represents the whole environment, while its observed and unobserved part are constituted just by sets of oscillators related to different frequency ranges, separated by a cut $\beta$. Here, the $fE$ is represented by the oscillators with frequency $\omega\in\left[0,\beta\right]$, while $(1-f)E$ is constituted by the complement 
$\omega\in\left[0,\beta\right]^c$. This situation corresponds to that sketched in Fig.\ \ref{SDCuts2} once one sets $\alpha=0$. 

\begin{figure}
\begin{center}
	
\includegraphics[width=0.8\columnwidth]{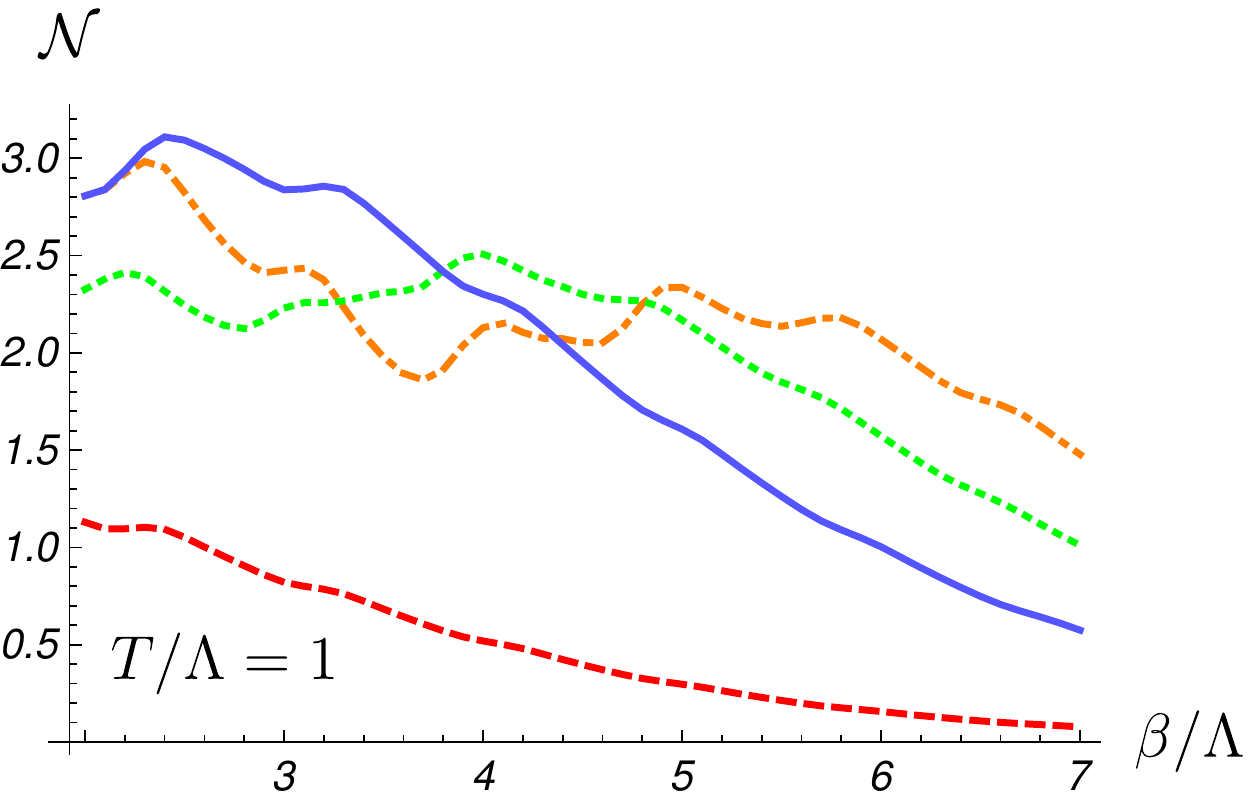}
\caption{\label{nalphaPlot1} Non-Markovianity measure in Eq.\ (\ref{NonMarkovMeasure}) as a function of $\beta$. 
The blue solid line and the red dashed one corresponds to the case in which there is only one cut in the spectral density and the observed frequencies belong to an interval $[0, \beta]$: the former is related to $s=4$, while the latter is related to $s=3$. 
The green dotted line and the orange dot-dashed one regards the situation where the observed environment is constituted by the oscillators with frequencies belonging to the interval $\left[\alpha,\beta\right]\equiv\left[\beta-\Delta,\beta\right]$, i.e., the so-called two cuts case. Here, the two lines are associated, respectively, to $\Delta=1$ and $2$, while both have been obtained for $s=4$. }
\end{center}

\end{figure}

An interesting question is how the measure in Eq.\ (\ref{NonMarkovMeasure}) depends on the cut $\beta$, distinguishing the observed and unobserved part of the environment. 
This behavior is presented in Fig.\ \ref{nalphaPlot1}. 
In general, the dependence on the cut does not show a monotonic trend, but there is a monotone decrease for large values of $\beta$, i.e. at the tight edge of the frequency domain.
The distinction between observed and unobserved environment may be implemented also in an another manner. We introduce two cuts, rather than just one. The observed part of the environment is constituted by the oscillators with frequency in the range $\omega\in\left[\alpha\equiv\beta-\Delta,\beta\right]$, while the unobserved one is related to the complementary set $\left[\alpha,\beta\right]^c$. This is sketched in Fig.\ \ref{SDCuts2}. We calculated the non-Markovianity measure as  a function of $\beta$ and various interval widths $\Delta$ for $s=4$ (Fig.\ \ref{nalphaPlot1}).  We see that the degree of non-Markovianity  depends on the position of the cuts in a non-monotonic way.

We remark on an important feature of the plot in  Fig.\ \ \ref{nalphaPlot1} for $s=3$  for the single cut case: The value of the measure is non-zero (see \cite{Haikka2013}). Thus, the introduction of the frequency cut  
changes the behavior from Markovian to non-Markovian. One can show it also for more than one cut. 
We conclude that the typical framework of quantum Darwinism, entailing a distinction into the observed and the unobserved parts of the environment, can lead to a certain amount of non-Markovianity, as certified by the measure (\ref{NonMarkovMeasure}).
We make one last remark concerning the implementation of the cuts. We used here sharp cuts defined by the step function. To exclude any artifacts connected with that, we also performed the parallel analysis using soft cuts
(modeled by the $\tanh  x$ function) and found that there is no quantitative difference between the two implementations. Thus, the non-Markovianity induced by the cuts does not depend on the their sharpness.

\section{Formation of Spectrum Broadcast Structures in  Non-Markovian environments}\label{DiscussionSec}
We now move to the main study of the current paper: the study of the connection between Markovianity and emergence of an objective value of the central spin in the sense of SBS formation.
We first study the full, uncut spectral density in Eq.\ (\ref{SD}) and then introduce the cuts, defining the observed and unobserved parts of the environment.
\subsection{Case I: Uncut spectral density}
In the first part of the discussion, we consider both observed and unobserved environment to be so large as to include the whole frequency domai; namely, we start with the simple case in which there are no cuts in the spectral density. 
In this case, the integrals in Eqs.\ (\ref{dec}) and (\ref{fid}), defining, respectively, the decoherence and the fidelity, are both performed  for $\omega\in[0,\infty)$, with the spectral density given by Eq.\ (\ref{SD})
-- note that $|\Gamma(t)|\leq B(t)$ because for $T>0$, $\tanh\left[\omega/(2T)\right]> \coth \left[\omega/(2T)\right]$. The integrals can be performed analytically. For the integer values of $s$, 
the results are presented below with both the decoherence and the fidelity factors decomposed into a vacuum and thermal parts: 
\begin{eqnarray}
\log |\Gamma(t)| = \log |\Gamma_{vac}(t)|+ \log |\Gamma_{th}(t)|,
\end{eqnarray}
where:
\begin{eqnarray}
\label{eq:decvac}
&& \frac{1}{2} \log |\Gamma_{vac}(t)| \equiv\wp(s-1)\left[1-\frac{\cos \left[(s-1)\arctan (\Lambda t)\right]}{\left(1+\Lambda^2 t^2\right)^{\frac{s-1}{2}}} \right]  \nonumber\\  &&\\
\label{eq:decth}
&&\frac{1}{2}\log |\Gamma_{th}(t)| \equiv \left(-\frac{T}{\Lambda}\right)^{s-1} \left[ 2\Psi^{(s-2)}\left(1+\frac{T}{\Lambda}\right) \right.- \nonumber \\  &&-\left.\Psi^{(s-2)}\left(1+\frac{T}{\Lambda}-iTt\right)  -c.c.\right],
\end{eqnarray}
and: 
\begin{equation}
\log B(t) = \log B_{vac}(t)+ \log B_{th}(t),
\end{equation}
where $B_{vac}(t)=|\Gamma_{vac}(t)|$ as for pure states decoherence and fidelity factors become equal, and:
\begin{eqnarray}
\label{eq:fidth}
&&\frac{1}{4}\log B_{th}(t) \equiv \left(-\frac{T}{2\Lambda}\right)^{s-1}\left\{\Psi^{(s-2)}\left(1+\frac{T}{2\Lambda}\right) -\nonumber \right. \\&& \left.-\Psi^{(s-2)}\left(\frac{1}{2}+\frac{T}{2\Lambda}\right)+\frac{1}{2}\Psi^{(s-2)}\left(\frac{1}{2}+\frac{T }{2\Lambda}+i\frac{Tt}{2}\right) \right. \nonumber  \\&& \left. -\frac{1}{2}\Psi^{(s-2)}\left(1+\frac{T}{2\Lambda}+i\frac{Tt}{2}\right) +c.c\right\}. 
\end{eqnarray} In the above formulas $c.c.$ stands for the complex conjugate, $\wp(z)$ is the Euler gamma function, and  $\Psi^{m}(z)$ is the so-called polygamma function defined as \cite{NIST}:
\begin{equation}
\Psi^{m}(z) \equiv \frac{d^{m+1}}{dz^{m+1}} \ln \wp (z) =  \sum_{k=0}^{\infty} \frac{(-1)^{m+1} m!}{(z+k)^{m+1}}.
\end{equation}
Generalization to the non-integer $s$ is presented in Appendix \ref{app:analytical}.


\begin{figure}
\begin{center}
\includegraphics[width=0.9\columnwidth]{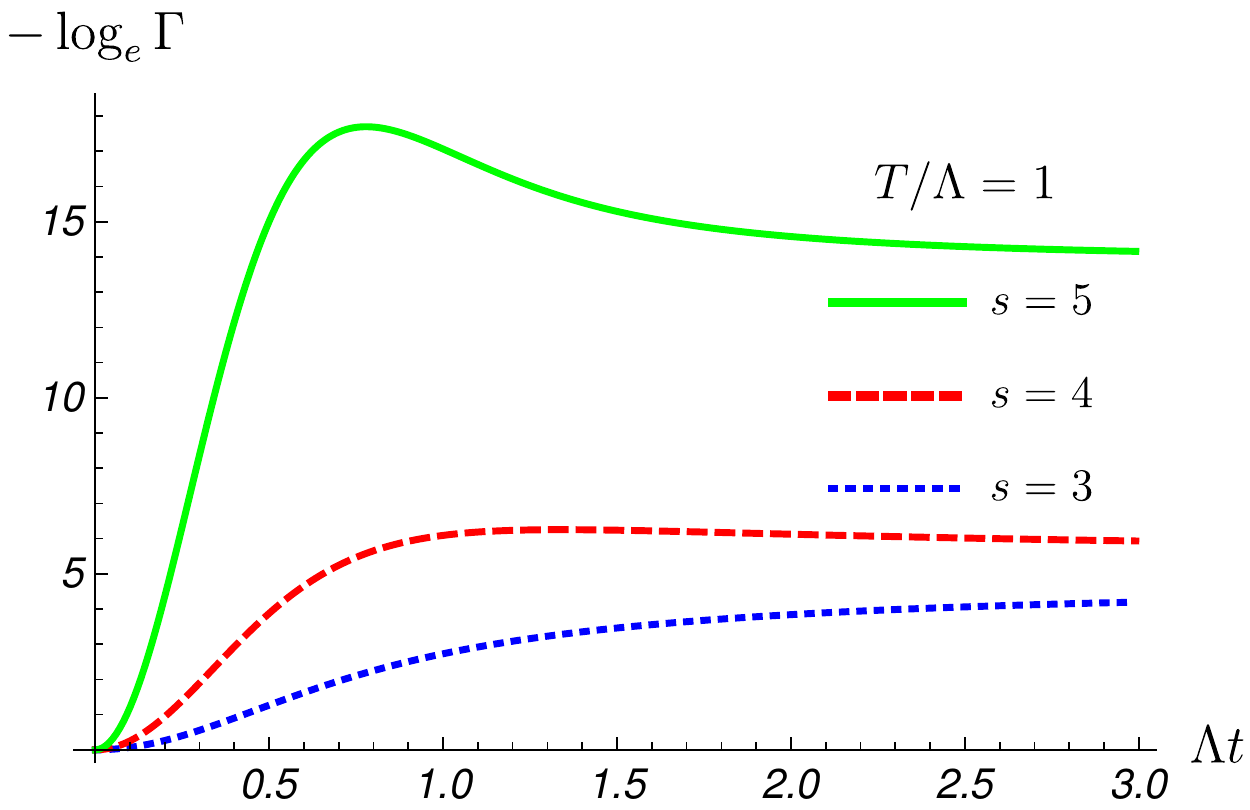}
\caption{\label{decWhole} Minus logarithm of the decoherence factor presented in Eq.\ (\ref{dec}) as a function of time for different
values of the $s$ parameter.}
\end{center}
\end{figure}
\begin{figure}
\begin{center}
\includegraphics[width=0.9\columnwidth]{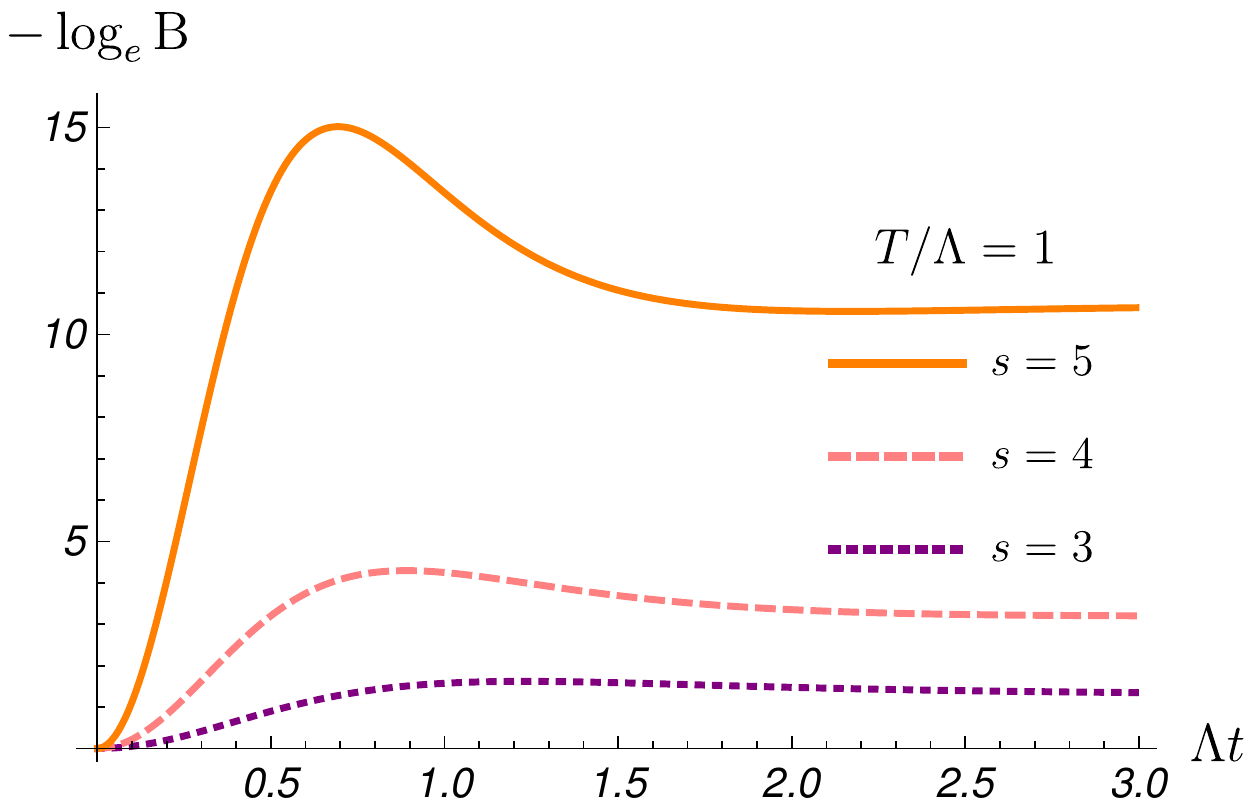}
\caption{\label{fidWhole} Minus logarithm of the fidelity presented in Eq.\ (\ref{fid}) as a function of the time for different values of the $s$ parameter.}
\end{center}
\end{figure}

From our point of view, the most interesting regime is the intermediate temperature one: 
\begin{equation}\label{iT}
0 \ll T \ll s \Lambda, 
\end{equation}
since
for low temperatures the decoherence and fidelity factors become identical modulo the macrofraction sizes. 
For high temperatures $T \gg s\Lambda$; on the other hand, by approximating the hyperbolic functions involved in the formulas one sees that the decoherence process is very strong but fidelity will be close to $1$, meaning that the environmental states become almost indistinguishable and hence no information about the state of the central system is stored in the environment -- it is too noisy.   
Unfortunately the above functions are too complicated for an analytical analysis for the temperature regime given by Eq.\ (\ref{iT}) and we resort here to a numerical analysis.

In Fig.\ \ref{decWhole} and \ref{fidWhole} we plot the minus logarithms of the indicator functions (the decoherence and the fidelity) for $T= \Lambda$. Their high values for times larger than the inverse cutoff 
indicate that the partially reduced state quite quickly approaches SBS.
We also see that the increase of the Ohmicity $s$ results in higher asymptotic values so that the higher $s$, the closer the partially reduced state is to SBS. 
Moreover, in Fig.\ \ref{nSPlot} we see that the degree of non-Markovianity also grows with $s$. Thus, by looking to the dependence on the Ohmicity parameter, 
non-Markovianity favors, rather than hinders, the emergence of SBS and, as a result, objectivity. 
One possible insight for that is that bigger $s$ corresponds to stronger coupling between the system and environment, as can be seen from Eq.\ (\ref{SD}).

\begin{figure}
\begin{center}
\includegraphics[width=0.9\columnwidth]{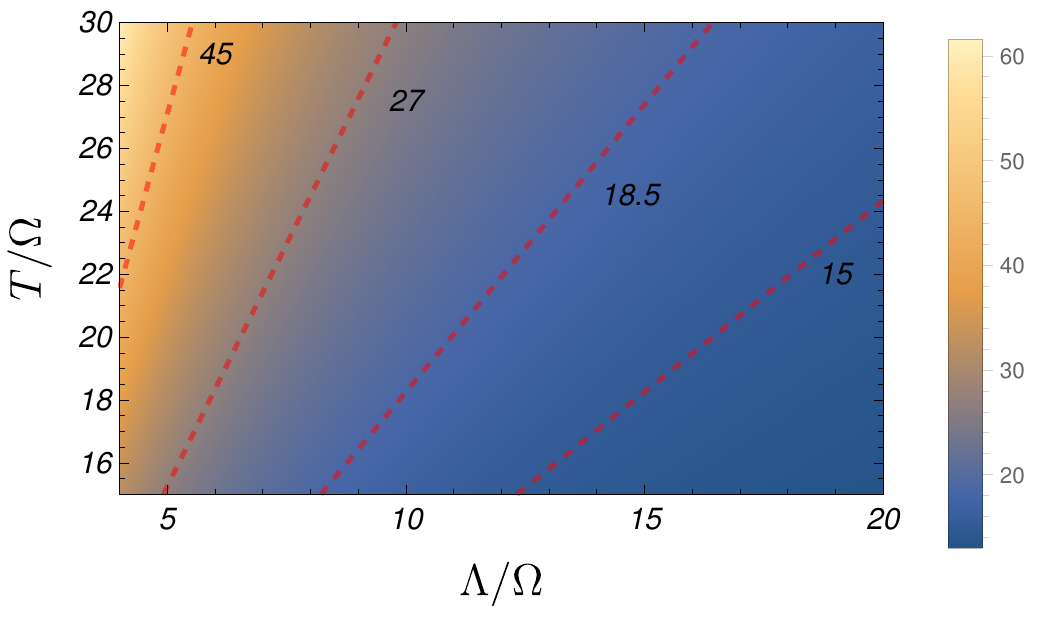}
\includegraphics[width=0.9\columnwidth]{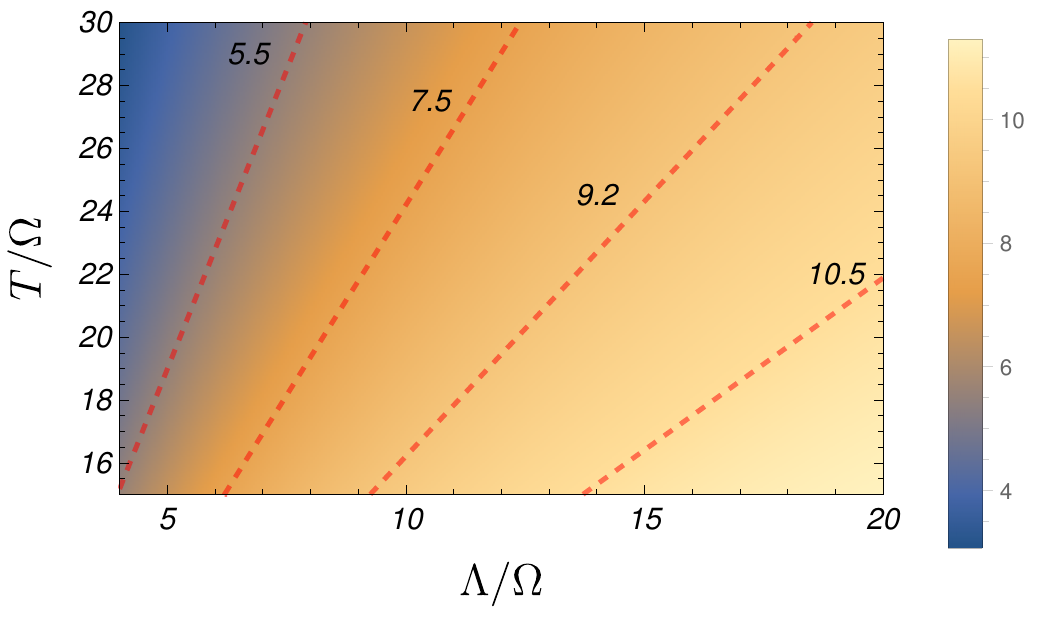}
\caption{\label{decTLambda} Asymptotic value of the minus logarithm of the decoherence factor presented in Eq.\ (\ref{dec}) (top) and of the fidelity presented in Eq.\ (\ref{fid}) (bottom) as a function of the temperature and the cutoff, at $s=5$.}
\end{center}
\end{figure}

The degree of non-Markovianity may be changed also by tuning other parameters, e.g. the cutoff $\Lambda$.
Fig.\ \ref{nSPlot} shows that non-Markovianity increases when $\Lambda \rightarrow 0$.
In Fig.\ \ref{decTLambda} we plotted minus logarithms of the asymptotic values of decoherence and fidelity, respectively, as a function of the cutoff and the temperature. 
We note that decoherence gets weaker as $\Lambda\rightarrow0$, i.e., when the non-Markovianity increases, while the fidelity shows the opposite behavior and in this case there seems to be no straightforward connection between the non-Markovianity 
and  the SBS formation.
%

The value of the non-Markovianity measure depends also on temperature $T$.
In Sec.\ \ref{NonMarkovianRegimeSec} we pointed out that in the high-temperature regime non-Markovianity decreases linearly with $T$, provided that $s>3$. However, the indicator functions depend on the temperature in opposite ways: 
With growing temperature decoherence gets stronger, while the states of the environment become harder to distinguish due to the higher thermal noise.
In this case there is also no clear connection between the degree of non-Markovianity and the formation of SBS.

Summarizing the uncut spectral density case,  when looking at the Ohmicity parameter $s$ alone, it seems that the stronger non-Markovianity enhances the formation of SBS. However, taking into account the other parameters $\Lambda, T$ does not support this claim. To the contrary, it seems that the non-Markovianity does not have a direct influence on the process of SBS formation in the considered model. The most important parameter is the Ohmicity $s$ that controls coupling strength between the system and the environment.     

\subsection{Case II: Cut spectral density}
\begin{figure}
\begin{center}
\includegraphics[width=1\columnwidth]{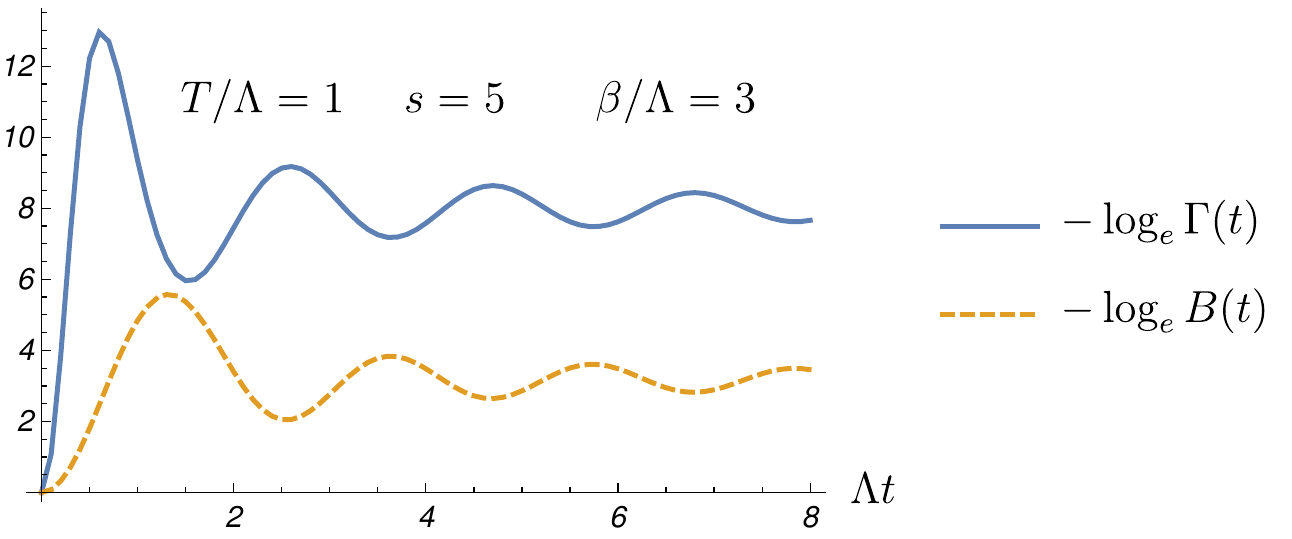}
\caption{\label{fidDecTime} Minus logarithm of the decoherence factor in Eq.\ (\ref{dec}) (blue solid line) and of the fidelity in Eq.\ (\ref{fid}) (orange dashed line) in the case when the observed frequencies are in the interval $\left[0, \frac{\beta}{\Lambda}\right]$.}
\end{center}
\end{figure}

We consider now the situation in which the whole environment is modeled through the spectral density in Eq.\ (\ref{SD}), while  the observed part $fE$ and the unobserved one $(1-f)E$ 
are represented only by its different fragments. We begin with a single cut case:  The observed and the unobserved environments are defined in the frequency domain by a single cut located at some frequency $\beta$; 
see Fig.\ \ref{SDCuts2} with $\alpha=0$. Unlike in the previous case of the uncut environment, the analytical solution is not possible and we have to resort to a numerical analysis straight from the beginning.
In Fig.\ \ref{fidDecTime} we plotted the minus logarithms of the decoherence factor and the fidelity as a function of time.
We note the presence of oscillations in the time evolution of both functions. These oscillations indicate a non-Markovian behavior since they constitute non-monotonicity areas. 
They occur even for $s=3$, when the behavior is pure Markovian for the uncut case \cite{Haikka2013}. This means that the presence of the cut turns on non-Markovianity effects, as we have already shown in Fig.\ \ref{nalphaPlot1}.  

\begin{figure}
\begin{center}
\includegraphics[width=1\columnwidth]{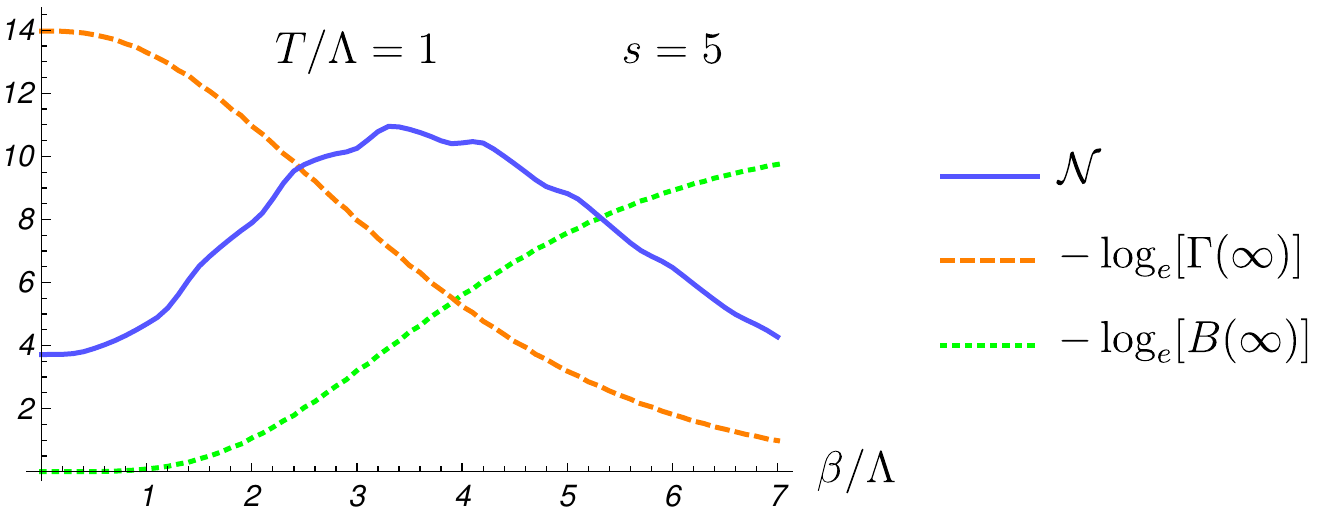}
\caption{\label{OneCutComparison} Asymptotic value of the minus logarithm of the decoherence factor in Eq.\ (\ref{dec}) (orange dashed line) and of the fidelity in Eq.\ (\ref{fid}) (green dotted line) in the case when the observed frequencies are in the interval $\left[0, \frac{\beta}{\Lambda}\right]$.
The blue solid line represents the value of the non-Markovianity measure in this situation as a function of the cut frequency. }
\end{center}
\end{figure}

Fig.\ \ref{fidDecTime} also shows that after a very long time ($\Lambda t\gg1$) both decoherence and fidelity approach constant values. In Fig.\ \ref{OneCutComparison} we plot these asymptotic values as functions of the frequency cut $\beta$.
This plot shows an interesting reciprocal behavior: While decoherence gets stronger when $\beta$ decreases, state distinguishability gets weaker. This is in agreement with the fact that decoherence is related to the unobserved environment while distinguishability is related to the observed one. When $\beta$ increases, the observed environment enlarges, so the state distinguishability gets better, while the unobserved environment shrinks which negatively affects the decoherence process.
In Fig.\ \ref{OneCutComparison} we also plot the value of the non-Markovianity measure as a function of the frequency cut $\beta$. In the lower part of the frequency domain, the growth of $\beta$ corresponds to both an increase of the non-Markovian effects
and an approach to SBS. 
Past the maximum at approximately $\beta/\Lambda =4$, both the non-Markovianity and the fidelity to SBS decrease (the latter due to the decreased decoherence). Thus, in this  particular case it is possible to link non-Markovianity with the approach to objectivity.

\begin{figure}
	\begin{center}
		\includegraphics[width=1\columnwidth]{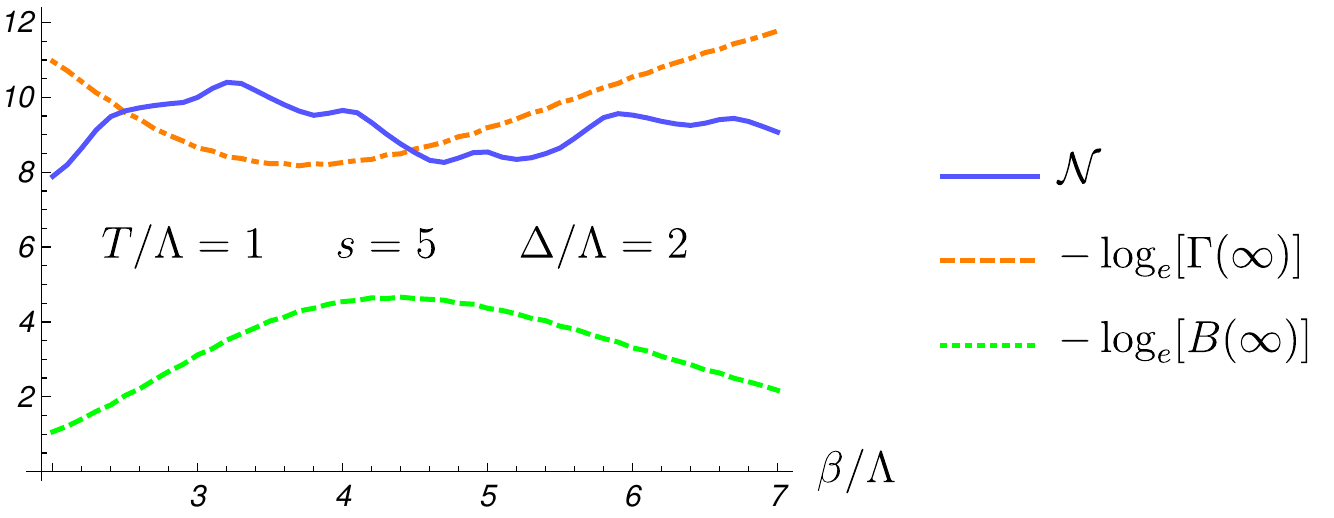}
		\caption{\label{twocutsdec} Asymptotic value of minus logarithm of the decoherence factor in Eq.\ (\ref{dec}) (orange dashed line) and of the fidelity in Eq.\ (\ref{fid}) (green dotted line) in the case when the unobserved fraction of the environment are in the complement of the interval $\left[ \frac{\alpha}{\Lambda},\frac{\beta}{\Lambda}\right]\equiv\left[ \frac{\beta-\Delta}{\Lambda},\frac{\beta}{\Lambda}\right]$, see also Fig.\ \ref{SDCuts2}. 
The blue solid line represents the value of the non-Markovianity measure in this situation as a function of the cut frequency $\beta$, when $\Delta/\Lambda=2$.		
		}
	\end{center}
\end{figure}

Finally, we investigate the two cuts case, i.e., when the observed part of the environment corresponds to a frequency window and the unobserved parts correspond to the rest of the spectrum.
The results, presented in Fig.\ \ref{twocutsdec}, show that moving the frequency window towards mid frequencies leads to decrease of decoherence and  increase of distinguishability. When the cuts are placed around  $ \alpha= 3\Lambda, \beta =5 \Lambda$, the partially reduced state is a good approximation to SBS.  On the other hand, shifting the frequency window towards the upper part of the frequency domain results in increase of decoherence and decrease of distinguishability. The value of the non-Markovianity measure, also plotted in Fig.\ \ref{twocutsdec},  is a complicated function of the frequency window location. Thus, in this case, there is no straightforward connection between non-Markovianity and the approach to objectivity.   

\section{Conclusions}
We studied the emergence of a SBS in a paradigmatic model of an open quantum system: the pure dephasing spin-boson model. 
SBS represents the structure of the quantum states which encode objective property after the interaction with the environment. 
We showed that this structure arises in the spin-boson model as a result of the temporal evolution.
In this case what becomes objective is the state of the central spin, which, after SBS formation, is perceived by many observers as a classical bit with values $\pm 1/2$.
This is an original result of our paper, which enforces the reliability of SBS to describe in terms of state the emergence of objectivity of classical theory starting by the underlying quantum domain. 

A large part of the paper has been devoted to the analysis of if and how the presence of non-Markovian effects influences the SBS formation. 
This task has been already investigated in \cite{Giorgi2015} for a qubit in a spin environment and \cite{Galve2016} in the context of quantum Brownian motion, using quantum Darwinism rather than a more fundamental SBS approach. 
We discuss a particular situation where non-Markovianity favors the formation of a SBS. 
This is the case in which we control non-Markovianity degree by tuning the degree of super Ohmicity. 
Our insight is that, rather than non-Markovianity, there are other physical properties that are decisive in the process of emergence of a SBS, for instance strength of  the coupling. 

We also showed that in the framework of quantum Darwinism, where environment is divided into observed and unobserved parts, there is a certain amount of non-Markovianity caused by this ``environment cutting".

The analysis of the emergence of the SBS for the present model opens also the possibility to study objectivity from the experimental point of view. 
In fact, the pure dephasing spin-boson model admits several practical counterparts. 
In \cite{Cirone2009, Haikka2011} it has been shown that an experimental realization of the spin-boson model may be obtained by means of an impurity in an ultracold gas. In this context the degrees of freedom of the gas play the role of the oscillators in Eq.\ (\ref{FreeHamiltonians}), while the two-level system may be constructed putting the impurity in a double-well trap potential.  
In \cite{Liu2011}, instead, a realization of the pure dephasing spin-boson model with photons has been presented: The two polarization states correspond the two-level open system while the bosonic modes of the environment are represented by the frequency degree of freedom of the photon which is coupled to the system
via an interaction induced by a birefringent material.

\acknowledgments 
Insightful discussion with Pawe\l{} Horodecki, Manabendra Nath Bera, Vincenzo D'Ambrosio, Arnau Riera, Jan Wehr, Bogna Bylicka and Dariusz Chru\'{s}ci\'{n}ski are  gratefully  acknowledged. 

This work has been funded by a scholarship from the Programa M\'{a}sters d'Excel-l\'{e}ncia of the Fundaci\'{o} Catalunya-La Pedrera, ERC Advanced Grant OSYRIS, EU IP SIQS, EU PRO QUIC, 
EU STREP EQuaM (FP7/2007-2013, No. 323714), Fundaci\'o Cellex, the Spanish MINECO (SEVERO OCHOA GRANT SEV-2015-0522,  FOQUS FIS2013-46768, FISICATEAMO FIS2016-79508-P), and the Generalitat de Catalunya (SGR 874 and CERCA/Program).
The work was made possible
through the support of grant from the John Templeton
Foundation (JT and JKK). The opinions expressed in this publication
are those of the authors and do not necessarily reflect
the views of the John Templeton Foundation. JT acknowledges
support of the Polish National Science Center by means of project no. 2015/16/T/ST2/00354 for the PhD thesis.


\appendix
\section{Canonical form of a time-local Master equation describing evolution of the partially reduced state}
	\label{app:canlk}
	In this appendix we show that evolution of the partially reduced state [Eq.\ (\ref{SBS2}] of the main text) can be rewritten in the canonical form of a time-local master equation \cite{Hall2014}:
	\begin{eqnarray}
	\label{eq:canL}
	&&\dot{\rho}_{S:fE} = -i \left[H(t),\rho_{S:fE}(t)\right]   \\ \nonumber &&+  \gamma(t) \left[L(t) \rho_{S:fE} L^{\dagger}(t)- \frac{1}{2}\left\lbrace L^{\dagger}(t)L(t), \rho_{S:fE} \right\rbrace \right], 
	\end{eqnarray} where for sake of the simplicity explicit time dependence of the partially reduced density matrix $\rho_{S:fE}$ was omitted and time derivation was denoted by $\dot{\rho}$.

Using Eq.\ (\ref{SBS2}) of the main text we can write:
	\begin{eqnarray}
	\dot{\rho}_{S:fE} = \begin{pmatrix}
	c^{00}_S(t) \rho^{00}_f(t) \; & c^{01}_S(t) \rho^{01}_f(t)  \\ & \\
	c^{10}_S(t) \rho^{10}_f(t)  \; & c^{11}_S(t) \rho^{11}_f(t) 
	\end{pmatrix},
	\end{eqnarray}
	where:
	\begin{eqnarray}
	&&c^{nm}_S(t) \equiv  \begin{cases}
	c^{nn}_{0S}       & \quad \text{for } n =m\\
	\Gamma(t)c^{nm}_{0S}  & \quad \text{for } n \neq m \\
	\end{cases} \\
	\end{eqnarray}
	with, real for initial thermal states, factor:
	\begin{eqnarray}
	&&\Gamma(t)=\prod_{k\in(1-f)E}\text{Tr}\left[U_{E_k}(n,t)\rho_{0k}U_{E_k}(m,t)^{\dagger}\right] \nonumber
	\end{eqnarray}
	and (as in the main text):
	\begin{eqnarray}
	&&\rho^{nm}_{f}(t)\equiv\bigotimes^{fN}_{k=1}U_{E_k}(n,t)\rho_{0k}U_{E_k}(m,t)^{\dagger}
	\end{eqnarray}
	The first step in derivation is calculation of $\dot{\rho}_{S:fE}$. We find:
	\begin{eqnarray}
	\label{eq:derivative}
	\dot{\rho}_{S:fE} = &&\begin{pmatrix}
	c^{00}_S(t) \dot{\rho}^{00}_f(t) \; & c^{01}_S(t) \dot{\rho}^{01}_f(t)  \\ & \\
	c^{10}_S(t) \dot{\rho}^{10}_f(t)  \; & c^{11}_S(t) \dot{\rho}^{11}_f(t) 
	\end{pmatrix} + \\ \nonumber  &&\gamma(t) \begin{pmatrix}  0  \; & c^{01}_S(t) \rho^{01}_f(t) \\  c^{10}_S(t) \rho^{10}_f(t)\;  &  0 
	\end{pmatrix},
	\end{eqnarray} 
	where (c.f. Eq.\ (\ref{eq:cham}) of the main text):
	\begin{eqnarray}
	&&\dot{\rho}^{nm}_f(t) = -i \bigotimes_{k=1}^{fN}\left(H^n_k(t) \rho_{0k} +  \rho_{0k} H^m_k(t) \right) \equiv \\ \nonumber && -i\left( H^n_{fE}(t) \rho_{0f} +  \rho_{0f} H^m_{fE}(t) \right),   \\
	&&\gamma(t) = \frac{\dot{\Gamma}(t)}{\Gamma(t)}
	\end{eqnarray}
	Finally, defining: 
	\begin{equation}
	H_{S:fE}(t) \equiv \sum_n  \ket{n}\bra{n} \otimes H^n_{fE}(t),
	\end{equation}
	we can recast Eq.\ (\ref{eq:derivative}) as:
	\begin{eqnarray}
	\dot{\rho}_{S:fE} = &&-i\left[H_{S:fE}(t),\rho_{S:fE}\right] + \\ &&\gamma(t) \left( \sigma_z \otimes I_{fE} \rho_{S:fE} \sigma_z \otimes I_{fE} - \rho_{S:fE}\right). \nonumber
	\end{eqnarray} 
	After identification $L(t) \equiv \sigma_z \otimes I_{fE}$ we arrive at Eq.\ (\ref{eq:canL}).
	Quantity $\gamma(t)$ is called the canonical decoherence rate. If this rate is positive at all times, then the evolution over any time interval is completely positive \cite{Rivas2010,Hall2014}. On this basis \cite{Hall2014} the following definition of non-Markovian evolution was introduced.

{\it Definition. A time-local master equation is Markovian, at some given time, if and only if the canonical decoherence rates are positive. Correspondingly, the evolution is non-Markovian if one or more of the canonical decoherence rates is strictly negative.  
}

According to the definition, the integral \cite{Hall2014}:
	\begin{equation}
	\mathcal{N}=-\int_{\gamma<0}\gamma(t)dt,
	\end{equation} can be used to measure the total amount of non-Markovianity for the considered evolution. 

\section{Fidelity in the spin-boson model}\label{FidelitySec}
In this appendix we present a detailed derivation of the expression for the fidelity. 
The quantity we want to evaluate is:  
\begin{equation}
B_{nm}(t)\equiv B\left[\rho^{(k)}_{nn}(t),\rho^{(k)}_{mm}(t)\right].
\end{equation}
denoting a single-subsystem overlap. 
Dropping the explicit dependence on the index $k$ we obtain:
\begin{equation}\label{Bmic}
B_{nm}(t)=\text{Tr}\sqrt{\sqrt{\rho_0}U(m;t)^\dagger U(n;t)\rho_0 U(n;t)^\dagger U(m;t)\sqrt{\rho_0}},\\
\end{equation}
where we have pulled the extreme left and right unitaries out of both the square roots and used the cyclic property of the trace to cancel them out.
The free evolutions $e^{-inE_kt}$ cancel out as both unitary 
operators under the square root are Hermitian conjugates of each other.
Thus, modulo phase factors:
\begin{align}
U(m;t)^\dagger U(n;t)\simeq D\left(\alpha(t)(n-m)\right)\equiv D(\eta_t),\label{et}
\end{align}
where $D(\alpha)$ represents the displacement operator. 
Next, assuming all the initial states $\rho_{0k}$ are thermal with the same temperature, we use the corresponding $P$-representation for the middle $\rho_0$ under the square root in (\ref{Bmic}):
\begin{equation}\label{ThermalInitialCnditions}
\rho_0=\rho_{th}(\bar{n})\equiv\frac{1}{\bar n}\int\frac{d^2\gamma}{\pi}e^{-\frac{|\gamma|^2}{\bar n}}\ket\gamma\bra\gamma, 
\end{equation}
where $\bar n=1/(e^{\beta \omega}-1)$, $\beta\equiv1/T$. 
Indicating the Hermitian operator under the square root in (\ref{Bmic}) by $\tilde B_t$, we obtain:
\begin{align}\label{B1}
\tilde{B}_t=&\int\frac{d^2\gamma}{\pi\bar n}e^{-\frac{|\gamma|^2}{\bar n}}\sqrt{\rho_0} D(\eta_t)\ket\gamma\bra\gamma D(\eta_t)^\dagger\sqrt{\rho_0}\nonumber\\
=&\int\frac{d^2\gamma}{\pi\bar n}e^{-\frac{|\gamma|^2}{\bar n}}\sqrt{\rho_0}\ket{\gamma+\eta_t}\bra{\gamma+\eta_t}\sqrt{\rho_0}.
\end{align}
The next step is to calculate  explicitly the square root in the equation above. For this aim we expand $\rho_0$ in the Fock basis:
\begin{equation}\label{FockRepresentation}
\rho_0=\sum_k\frac{\bar n^k}{(\bar k +1)^{n+1}}\ket k\bra k.
\end{equation}
Replacing it in Eq.\ (\ref{B1}) we have:
\begin{equation}\label{B2}
\tilde{B}_t=\int\frac{d^2\gamma}{\pi\bar n}e^{-\frac{|\gamma|^2}{\bar n}}\sum_{i,j}\lambda_{ij}(\bar n)
{\langle j|\gamma+\eta_t\rangle}{\langle \gamma+\eta_t|i\rangle}\ket{j}\bra{i}
\end{equation}
with:
\begin{equation}
\lambda_{ij}(\bar n)\equiv\sqrt{\frac{\bar n^{i+j}}{(\bar n +1)^{i+j+2}}}.
\end{equation}
Using the Fock basis $\ket{j}$ representation of coherent states one may get the explicit scalar product $\langle j|\gamma+\eta_t\rangle$. Accordingly Eq.\ (\ref{B2}) gets:
\begin{align}
\tilde{B}_t&=\frac{1}{\bar n +1}e^{-\frac{|\eta_t|^2}{1+2\bar n}}\int\frac{d^2\gamma}{\pi\bar n}e^{-\frac{1+2\bar n}{\bar n(\bar n +1)}
\left|\gamma+\frac{\bar n}{1+2\bar n}\eta_t\right|^2}\times\nonumber\\
&\times \left|\sqrt{\frac{\bar n}{\bar n +1}}(\gamma+\eta_t)\right\rangle\left\langle\sqrt{\frac{\bar n}{\bar n +1}}(\gamma+\eta_t)\right|.\label{Btilde}
\end{align}
We now show that this equation is formally equivalent to that of a thermal state introduced in Eq.\ (\ref{ThermalInitialCnditions}). 
For this aim, we underline that we are interested in the square root of the operator  $\tilde{B}_t$, rather than in itself. Therefore, there is a freedom of rotating $\tilde{B}_t$ by a unitary operator, and in particular a displacement one:
\begin{equation}\label{DroppingDisplacement}
\text{Tr}\left[\sqrt{D\tilde{B}_tD^\dagger}\right]=
\text{Tr}\left[D\sqrt{\tilde{B}_t}D^\dagger\right]=
\text{Tr}\left[\sqrt{\tilde{B}_t}\right].
\end{equation}
In particular we find:
\begin{align}\label{DisplacementShift}
&\ket{\sqrt{\frac{\bar n}{\bar n +1}}(\gamma+\eta_t)}\varpropto\nonumber\\ 
&D\left(\sqrt{\frac{\bar n}{\bar n +1}}\frac{1+\bar n}{1+2\bar n}\right)\ket{\sqrt{\frac{\bar n}{\bar n +1}}(\gamma+\frac{\bar{n}}{1+2\bar{n}}\eta_t)},
\end{align}
where we have omitted the irrelevant phase factor arising from the action of the displacement. 
We replace Eq.\ (\ref{DisplacementShift}) with Eq.\ (\ref{Btilde}). Dropping displacement operators due to Eq.\ ($\ref{DroppingDisplacement}$) and introducing the variable:
\begin{equation}
\tilde{\gamma}=\sqrt{\frac{\bar n}{\bar n +1}}\left(\gamma+\frac{\bar n}{1+2\bar n}\eta_t\right)
\end{equation}
one obtains:
\begin{equation}
B_{nm}(t)=\frac{e^{-\frac{1}{2}\frac{|\eta_t|^2}{1+2\bar n}}}{\sqrt{1+2\bar n}}\text{Tr}\sqrt{\rho_{th}\left(\frac{\bar n^2}{1+2\bar n}\right)}.
\end{equation}
In order to calculate  explicitly  the square root we recall the Fock expansion in Eq.\ (\ref{FockRepresentation}).
Finally, we get:
\begin{equation}
B_{nm}(t)=\exp\left[-\frac{(n-m)^2}{2}|\alpha_k(t)|^2\tanh\left(\frac{\beta \omega_k}{2}\right)\right], 
\end{equation}
and generalize it to the  fidelity over all macrofractions:
\begin{equation}
B^{\mathcal{M}}_{nm}(t)=\exp\left[-\frac{(n-m)^2}{2}\sum_{k\in\mathcal{M}}|\alpha_k(t)|^2\tanh\left(\frac{\beta \omega_k}{2}\right)\right].
\end{equation}
The expression in Eq.\ (\ref{fid}) follows by taking the limit of continuum spectrum in the above equation. 
\section{Analytical expressions for decoherence factor and fidelity} \label{app:analytical}

When both observed and unobserved fragments of environment can be described in terms of the full spectral density, the decoherence factor and fidelity of the environmental sates are given by:
\begin{eqnarray}
\label{eq:b1}
&&\log |\Gamma(t)| =  \\ &&\frac{2}{\Lambda^{s-1}}\int_0^{\infty} d\omega \omega^{s-2}e^{-\omega/\Lambda}\left[1-\cos\left(\omega t\right)\right]\coth\left(\omega/2T\right), \nonumber \\
\label{eq:b2}
&&\log B(t) = \\  && \frac{2}{\Lambda^{s-1}}\int_0^{\infty} d\omega \omega^{s-2}e^{-\omega/\Lambda}\left[1-\cos\left(\omega t\right)\right]\tanh\left(\omega/2T\right). \nonumber 
\end{eqnarray}
In what follows we assume that $s$ is an integer number such that $s>1$ (the case for s=1 needs to be treated separately). Integrals in Eqs.\ (\ref{eq:b1}) and (\ref{eq:b2}) can be expressed in terms of Hurwitz zeta function \cite{NIST}:
\begin{equation}
\zeta(z,q) = \sum_{n=0}^\infty \frac{1}{(q+n)^z},
\end{equation}
what leads to the following expressions for decoherence factor: 
\begin{eqnarray}
&&\frac{1}{2}\log |\Gamma(t)| =  \wp(s-1)\left(\frac{T}{\Lambda}\right)^{s-1} \left[ 2\zeta\left(s-1,\frac{T}{\Lambda}+1\right) \right.- \nonumber \\  &&-\left.\zeta\left(s-1,\frac{T}{\Lambda}+1-iTt\right)  -c.c.\right],
\end{eqnarray}
where $\wp(s-1)$ is the Euler gamma function and  $c.c$ denotes complex conjugate. Similarly we find:
\begin{eqnarray}
&&\frac{1}{4}\log B(t) = \wp(s-1) \left(\frac{T}{2\Lambda}\right)^{s-1}\left[\zeta\left(s-1,1+\frac{T}{2\Lambda}\right) -\nonumber \right. \\&& \left.-\zeta\left(s-1,\frac{1}{2}+\frac{T}{2\Lambda}\right)+\frac{1}{2}\zeta\left(s-1,\frac{1}{2}+\frac{T }{2\Lambda}+i\frac{Tt}{2}\right) \right. \nonumber  \\&& \left. -\frac{1}{2}\zeta\left(s-1,1+\frac{T}{2\Lambda}+i\frac{Tt}{2}\right) +c.c.\right]. \nonumber
\end{eqnarray}
For integer $s \geq 2$, using the following relation between Hurwitz theta function and polygamma function \cite{NIST}:
\begin{equation}
\Psi^{m}(z)= (-1)^{m+1} \wp(m+1)\zeta \left(m+1,z\right),
\end{equation}
one reach Eqs.\ (\ref{eq:decth},\ref{eq:fidth}) presented in the main text.

\bibliography{Objectivity}

\end{document}